
\headline={\ifnum\pageno=1\firstheadline\else
\ifodd\pageno\rightheadline \else\leftheadline\fi\fi}
\def\firstheadline{\hfil}
\def\rightheadline{\hfil}
\def\leftheadline{\hfil}
        \footline={\ifnum\pageno=1\firstfootline\else\otherfootline\fi}
\def\firstfootline{\rm\hss\folio\hss}
\def\otherfootline{\hfil}
\font\tenbf=cmbx10
\font\tenrm=cmr10
\font\tenit=cmti10
\font\elevenbf=cmbx10 scaled\magstep 1
\font\elevenrm=cmr10 scaled\magstep 1
\font\elevenit=cmti10 scaled\magstep 1

\nopagenumbers
\line{\hfil }
\vglue 1cm
\hsize=6.0truein
\vsize=8.5truein
\parindent=3pc
\baselineskip=10pt

\def \po {$|\psi _0\rangle$}

\def \u {\vert \uparrow }

\def \d {\vert \downarrow }
\def \n {{1\over \sqrt 2}}

hep-th\9306086
\centerline{\tenbf  MEASUREMENT OF NONLOCAL VARIABLES}
\vglue 0.2cm
\centerline{\tenbf WITHOUT BREAKING CAUSALITY}

\vglue 1.0cm
\centerline{\tenrm LEV VAIDMAN }
\baselineskip=13pt
\centerline{\tenit
School of Physics and Astronomy}
\baselineskip=12pt
\centerline{\tenit
Raymond and Beverly Sackler Faculty of Exact Sciences}
\baselineskip=12pt
\centerline{\tenit
Tel-Aviv University, Tel-Aviv, 69978 ISRAEL}

\vglue 0.8cm
\centerline{\tenrm ABSTRACT}
\vglue 0.3cm
{\rightskip=3pc
 \leftskip=3pc
 \tenrm\baselineskip=12pt
 \noindent
We report results of an investigation of relativistic causality
constraints on the measurability of nonlocal variables. We show that
measurability of certain nondegenerate variables with entangled
eigenstates
contradicts the principle of causality, but that there are other,
certainly nonlocal, variables which can be measured without breaking
causality. We show that any causal measurement of nonlocal variables
must erase certain local information. For example, for a system of two
spin-1/2 particles, even if we take the weakest possible definition of
verification measurement, verification of an entangled state must
erase all local information.

\vglue 0.8cm }
\line{\elevenbf 1. Measuring momentum of a particle \hfil}
\vglue 0.4cm
\baselineskip=14pt
\elevenrm

As early as 1931, Landau and Peierls$ \vphantom{T}^1$ showed that
relativistic causality imposes new restrictions on the process of
quantum measurement.  Although some of their arguments were not
precise, it was commonly accepted that we cannot measure
instantaneously nonlocal properties without breaking relativistic
causality.

The first example is the measurement of momentum of a particle.
Consider a particle localized in a small region.  Measurement of its
momentum, irrespective of the outcome, will spread the particle all
over the space.  There will be a non-zero probability to find the
particle at a very large distance from its original place immediately
after the (instantaneous) momentum measurement, so it seems that the
particle moves faster than light.  However, this argument is not
decisive.  Relativistic causality states that it is impossible to send
a {\elevenit signal} with superluminal velocity.  It does not forbid
instantaneous measurement of momentum, say at $t=0$.  The
instantaneous measurement interaction will take place all over the
space and it can create particles everywhere.  Thus, the probability
of finding the particle at a given location after the momentum
measurement might be independent of what we did to the particle
located far away before the measurement.  Therefore, the possibility
of instantaneous momentum measurement does not lead automatically to
the possibility of sending signals with superluminal velocity.

Nevertheless, if we can measure the momentum of a spin-1/2 particle without
affecting its spin, then we can violate causality.
Indeed, let us assume that we know that at time $t=0$ the
momentum measurement will be performed.  At the time $t=-\epsilon$ we decide to
prepare the state of the particle ``up" or ``down" according to the
signal we want to send.  Then we can measure the spin component of the
particle which is detected at time $t=+\epsilon$ far from its
original location and thus send information with
superluminal velocity.  (The probability of finding the particle at a
given place is very small, but we can use a large ensemble of
identical particles and thus we can build a reliable superluminal
transmitter.)

\vglue 0.6cm
\line{\elevenbf 2. Constraints on Nonlocal Measurements of
Two Spin-1/2 Particles \hfil} \vglue
0.4cm

Although momentum measurement is a basic problem, it is still not the
simplest example we may consider.  Significant progress in
understanding causality constraints on quantum measurement was made by
considering an even simpler example: measurements of spin variables of
two spin-1/2 particles separated in space.  This is the system on
which Bohm and Aharonov$\vphantom{T}^2$ and later Bell$^3$ analyzed
the EPR argument and reached far-reaching conclusions regarding the
nonlocal structure of quantum theory.

In order to show how measurability of nonlocal variables contradicts
relativistic causality let us consider an operator with the following
nondegenerate eigenstates:

$$\eqalign{|\psi_1\rangle =& \u  \rangle_{_1} \u
\rangle_{{\vphantom A}_2}\cr
|\psi_2\rangle =& \d \rangle_1 \d \rangle_2\cr
|\psi_3\rangle =&\n \bigl(\u  \rangle_1 \d \rangle_2+
\d  \rangle_1 \u \rangle_2
\bigr)\cr
|\psi_4\rangle =&\n \bigl(\u  \rangle_1 \d \rangle_2-
\d  \rangle_1 \u \rangle_2
\bigr)\cr}\eqno(1)$$

\noindent This operator corresponds to a nonlocal variable because its
eigenstates are nonlocal.  We call the state of the composite system
nonlocal when it cannot be represented as a product of states
corresponding to localized parts of the system; these states are also
known as  {\elevenit entangled} states.

     Let us show that the measurability of this variable contradicts
relativistic causality. To this end we perform the following set of
measurements:

i) We prepare state $\u \rangle_2$ of particle number 2 a long
time before the time $t=0$.

ii) At time $t= -\epsilon$ we prepare state $\u \rangle_1$ or
$\d \rangle_1$
of particle number 1 according to the message we want to
send from particle 1 to particle 2.

iii) At time $t=0$ we measure the variable
defined by the nondegenerate eigenstates of Eq. (1).

 iv) At the time $t= \epsilon$ we measure the spin component of particle 2.

\noindent
The two events, choosing the spin of particle 1 and measurement of the
spin of particle 2, are space-like separated, and therefore must be
causally disconnected.  But if we choose spin ``up" for particle 1,
then the state of the composite system before the time $t=0$ is $ \u
\rangle_{_1} \u \rangle_{{\vphantom A}_2}$, the measurement at the
time $t=0$ does not change it (since it is an eigenstate), and thus
the spin measurement of particle two will yield ``up" with probability
one.  If, instead, at the time $t= -\epsilon$, we put, the particle 1
in the state ``down\rlap,"\ then the state of the composite system
before the measurement (iii) is $ \d \rangle_{_1} \u
\rangle_{{\vphantom A}_2}$.  This state is not one of the eigenstates
of the nonlocal operator, and therefore the measurement at time $t=0$
will change it.  Since the scalar product between $ \d \rangle_{_1} \u
\rangle_{{\vphantom A}_2}$ and the eigenstates is not vanishing only
for the eigenstates $|\psi_3\rangle$ and $|\psi_4\rangle$, the state
after $t=0$ will be one of those.  But for both $|\psi_3\rangle$ and
$|\psi_4\rangle$ the probability to find the spin ``up" for particle 2
is just 1/2.  We have shown that the possibility of measuring nonlocal
variable described by eigenstates (1) allows us to change the
probability of the result of a spin measurement performed on particle
2 by acting on particle 1 a time only $2\epsilon$ before the
measurement on particle 2; and since the distance between the
particles might be larger than $2\epsilon c$, this procedure
represents a superluminal signal transmitter.

\vglue 0.6cm \line{\elevenbf 3. Measurable Nonlocal Variables \hfil}
\vglue 0.4cm The examples above may lead us to believe that
measurement of any nonlocal variable breaks relativistic causality.
This, in fact, was generally believed until Aharonov and
Albert$\vphantom{T}^4$
found a method involving solely local interactions (hence consistent
with the causality principle) which does allow us to measure certain
nonlocal variables.  In particular, we can measure the variable
${\sigma_1}_z + {\sigma_2}_z $. The method applies the standard von
Neumann measuring procedure to a measuring device consisting of two
parts which were prepared in an {\elevenit entangled} state before the
measurement.  Each part of the measuring device interacts with one of
the particles for a short time, and is observed immediately after by a
local observer.  The combined observations of the two observers (one
at each particle) determines whether the state is $|\psi_1\rangle$,
$|\psi_2\rangle$ or belongs to the subspace spanned by
$|\psi_3\rangle$ and $|\psi_4\rangle$.  The feature of this
method
is that while it measures ${\sigma_1}_z + {\sigma_2}_z = 0$, it does
{\elevenit not} measure the spin of each particle separately.  The
details of the method of nonlocal measurements can be found in
Ref. (5).

It might seem that the measurability of the operator ${\sigma_1}_z +
{\sigma_2}_z$ has something to do with its having a complete set of
eigenstates which are not entangled.  But this is not the explanation.
The next example shows an operator with nondegenerate eigenstates that
are all entangled but which is, nevertheless, measurable by local
interactions.  The eigenstates of the nondegenerate operator are

$$\eqalign{
|\psi_1\rangle =&\n \bigl(\u  \rangle_1 \u \rangle_2+
\d  \rangle_1 \d \rangle_2
\bigr)\cr
|\psi_4\rangle =&\n \bigl(\u  \rangle_1 \u \rangle_2-
\d  \rangle_1 \d \rangle_2
\bigr)\cr
|\psi_3\rangle =&\n \bigl(\u  \rangle_1 \d \rangle_2+
\d  \rangle_1 \u \rangle_2
\bigr)\cr
|\psi_4\rangle =&\n \bigl(\u  \rangle_1 \d \rangle_2-
\d  \rangle_1 \u \rangle_2
\bigr)\cr}\eqno(2)$$

\noindent
This operator can be measured$\vphantom{T}^6$ using a set of
 nonlocal operators with degenerate eigenstates (such
as ${\sigma_1}_z + {\sigma_2}_z$), where the particles 1 and 2 are
far from one another.
Recently\rlap,$\vphantom{T}^7$\ the measurability of
operators for two spin-1/2 particles has been analyzed, and
it was shown that the only
measurable nondegenerate operators are those with
eigenstates of two possible types:

$$\eqalign{|\psi_1\rangle =& \u _{z} \rangle_{_1} \u
_{z'}\rangle_{{\vphantom A}_2}\cr
|\psi_2\rangle =& \u _{z} \rangle_1 \d _{z'}\rangle_2\cr
|\psi_3\rangle =& \d _{z} \rangle_1 \u _{z'}\rangle_2\cr
|\psi_4\rangle =& \d _{z} \rangle_1 \d _{z'}\rangle_2\cr}\eqno(3a)$$

or

$$\eqalign{|\psi_1\rangle =&\n \bigl(\u _{z} \rangle_1 \u _{z'}\rangle_2+
\d _{z} \rangle_1 \d _{z'}\rangle_2
\bigr)\cr
|\psi_2\rangle =&\n \bigl(\u _{z} \rangle_1 \u _{z'}\rangle_2-
\d _{z} \rangle_1 \d _{z'}\rangle_2
\bigr)\cr
|\psi_3\rangle =&\n \bigl(\u _{z} \rangle_1 \d _{z'}\rangle_2+
\d _{z} \rangle_1 \u _{z'}\rangle_2
\bigr)\cr
|\psi_4\rangle =&\n \bigl(\u _{z} \rangle_1 \d _{z'}\rangle_2-
\d _{z} \rangle_1 \u _{z'}\rangle_2
\bigr)\cr}\eqno(3b)$$

\noindent
with spin polarized ``up" or ``down" along
directions $z$ and $z'$.

Operators of type (3a), although they refer to two separated spins,
are effectively local.  They can be measured simply by measuring the
$z$ component of spin of the first particle and the $z'$ component of
spin of the second particle.  Operators with the eigenstates (3b) are
truly nonlocal.  They can be measured$ \vphantom{T}^7$ in the same
way as an operator with eigenstates given in Eq.  (2) (a particular
case of Eq.  (3b)).

On the other hand\rlap,$\vphantom{T}^7$\ measurability of any
nondegenerate
operator with eigenstates not equivalent to
the forms (3a)
or (3b) implies the  possibility of superluminal
communication, i.e., violation of relativistic causality.

\vglue 0.6cm
\line{\elevenbf 4. State Verification Measurements \hfil}
\vglue 0.4cm

A measurement of a nondegenerate operator is also a state verification
measurement for all its eigenstates.  The weakest possible definition
of a state verification measurement which requires only {\elevenit
reliability} of the measurement is: the verification measurements of
the state $|\psi _0\rangle$ must always yield the answer ``yes" if the
measured system has the initial state $|\psi _0\rangle$, and must
always yield ``no" if the system is initially in an orthogonal state.
One may suspect that the verification of a state with canonical form
(Schmidt decomposition) different from
$$\n \bigl(\u _{z} \rangle_1 \u _{z'}\rangle_2+
\d _{z} \rangle_1 \d _{z'}\rangle_2
\bigr)\eqno(4)$$
(the form of the eigenstates in (3b)) contradicts relativistic
causality; i.e., that verification of a state
$$|\psi_1\rangle = \alpha \u _{z} \rangle_1 \u _{z'}\rangle_2 +
\beta \d _{z} \rangle_1 \d _{z'}\rangle_2 ,~~~~
\vert \alpha \vert \ne \vert \beta \vert \ne 0
 \eqno(5) $$
allows superluminal communication.  Indeed, it has been
 shown$\vphantom{T}^6$
that the type of measurements of entangled states described above,
i.e. nondemolition operator measurements with solely local
interactions, cannot measure the state given by the form (5).

However, an unmeasurable quantity should not represent physical
reality.  If we want to consider the quantum state as a physical
(versus
purely mathematical) concept, it must be measurable.  We do know how
to {\elevenit prepare} this state (the preparation procedure is also
frequently called measurement).  But the state (5) can also be
measured using a new type of verification measurement named an {\elevenit
exchange} measurement\rlap.$\vphantom{T}^6$\
The idea is
to make simultaneous short local interactions with parts of the
measuring device such that the states of the system and the measuring
device will be exchanged.  The novel point in this method is that {\elevenit
local} interactions exchange {\elevenit nonlocal} states.  The
result of the measurement  cannot be read by two
local observers;
we must bring the two parts of the measuring device to one place.  In
addition, this procedure has another unconventional property.  The
final state of the system is completely independent of its initial
state: it is just the initial state of the measuring device.  The
state of the system {\elevenit is completely erased} by this state
verification measurement.

 It has recently been proven$\vphantom{T}^7$ that {\elevenit any}
  verification  of the state

$$|\psi_1\rangle = \alpha \u _{z} \rangle_1 \u _{z'}\rangle_2 + \beta
\d _{z} \rangle_1 \d _{z'}\rangle_2 ,~~~~ \alpha , \beta \ne 0
\eqno(6) $$

\noindent
erases all local information.
 The probable
 outcome of a local spin measurement performed after the state
verification measurement is independent of the state of the composite
system
prior to the state verification. The example considered above of
a measurable
nondegenerate operator (2) trivially fulfills this result:
for all
eigenstates we have the property that the probability for any outcome
of local spin measurement is the same. There is no local information after
this nonlocal measurement.

\vglue 0.6cm
\line{\elevenbf 5. Conclusions \hfil}
\vglue 0.4cm

Let us formulate the last result for the somewhat more general case of
a system of two separated particles with several
 orthogonal states.
Consider the Schmidt decomposition of a state \po ~of this composite
system:

$$|\psi _0\rangle=\sum_i\alpha _i|i\rangle_1|i\rangle_2.\eqno(7)$$

\noindent
Here
${{|i\rangle_1}}$ and ${{|i\rangle_2}}$
are local orthonormal bases of states of the two particles.
Let us denote by $H^{(1)}$ and $H^{(2)}$ the Hilbert spaces of part 1
and part 2
respectively, and by $H^{(1)}_0$ and $H^{(2)}_0$ the subspaces of
$H^{(1)}$ and
$H^{(2)}$ which are spanned by the base vectors $|i\rangle_1$ and
$|i\rangle_2$
corresponding to coefficients $\alpha _i\neq 0$.
Then for all
initial states  which belong to the
Hilbert space $H^{(1)}_0\!\otimes H^{(2)}_{\vphantom 0}$, the
probabilities $p(\psi)$ for results of local measurements in part 1,
performed after
verification of the state \po, have no dependence on the initial
state.

Thus, the erasing effect of the proposed ``exchange" measurements is a
gene-ric property of any reliable, causal state verification
measurement.  The full implications of this result are not yet clear.
It already has helped complete the analysis of measurability of
nondegenerate operators discussed above.  It also has been used
to show$\vphantom{T}^7$ that
measurability of certain {\elevenit ideal measurements of the first kind}
contradicts relativistic causality, thus  placing a serious doubt
concerning
the possibility of generalizing axiomatic quantum theory to the
relativistic domain.

We would like to conclude by stressing the importance of measuring
nonlocal properties via local interactions (with separate parts of the
measuring device prepared in an entangled state).  The same method can
be used for so-called ``multiple-time" measurements$\vphantom{T}^8$
which open the way to many new quantum phenomena\rlap.$\vphantom{T}^9$

\vglue 0.6cm
\line{\elevenbf Acknowledgements \hfil}
\vglue 0.4cm
 The research was supported by grant 425/91-1 of the the Basic
Research Foundation (administered by the Israel Academy of Sciences and
Humanities).

\vglue 0.6cm
\line{\elevenbf  References \hfil}
\vglue 0.4cm
\item{1.}
L. Landau and R.~Peierls, {\elevenit Z. Physik} {\elevenbf 69} (1931)
56.

\item{2.}  D. Bohm and Y. Aharonov, {\elevenit  Phys. Rev.}
{\elevenbf
108}  (1957) 1070.

\item{3.}  J.~S.~Bell, {\elevenit  Physics} {\elevenbf 1} (64) 195.

\item{4.}  Y.~Aharonov and D.~Albert,
{\elevenit  Phys.  Rev.} {\elevenbf D21} (1980) 3316.

\item{5.}
Y.~Aharonov and D.~Albert,
{\elevenit  Phys.  Rev.} {\elevenbf D24} (1981) 359.

\item{6.} Y.~Aharonov, D.~Albert, and L.~Vaidman,
{\elevenit Phys.  Rev.}  {\elevenbf D34} (1986) 1805.

\item{7.} S. Popescu and L. Vaidman,
``Causality restrictions on nonlocal quantum measurements", preprint
of Tel-Aviv University TAUP-2011-92 (1992).

\item{8.}
Y.~Aharonov and D.~Albert,
{\elevenit  Phys.  Rev.} {\elevenbf D29} (1984) 223.

\item{9.}
 Y. Aharonov and L. Vaidman, {\elevenit J.Phys.  A: Math.  Gen.} {\elevenbf 24}
(1991) 2315.

   \bye